\documentclass[12pt]{article}

\usepackage{amsmath}
\usepackage{amsfonts}
\usepackage{amssymb}
\usepackage{graphics,graphicx,tikz}
\usepackage{soul}
\usepackage{cite}
\usetikzlibrary{patterns}

\numberwithin{equation}{section} 
\usepackage[linktocpage]{hyperref}
\hypersetup{
	colorlinks=true,
	linkcolor=blue,
	filecolor=magenta,      
	urlcolor=blue,
	citecolor=blue
}
\def\beq{\begin{equation}}
\def\eeq{\end{equation}}
\newcommand{\commentOut}[1]{}
\def\bea{\begin{align}}
\def\eea{\end{align}}

\usepackage{color}

\begin{document}
\begin{titlepage}
\hfill \hbox{NORDITA-2022-018}
\vskip 0.1cm
\hfill \hbox{UPPSALA-19/22}
\vskip 1.5cm
\begin{flushright}
\end{flushright}
\vskip 1.0cm
\begin{center}
{\Large \bf Radiation reaction for spinning black-hole scattering}
\vskip 1.0cm {\large  Francesco Alessio$^{a,b}$  and Paolo Di Vecchia$^{a, c}$  } \\[0.7cm]

{\it \small $^a$ NORDITA, KTH Royal Institute of Technology and Stockholm University, \\
 Hannes Alfv{\'{e}}ns v{\"{a}}g 12, SE-11419 Stockholm, Sweden  }\\
 {\it \small $^b$ Department of Physics and Astronomy, Uppsala University,\\ Box 516, SE-75120 Uppsala, Sweden}

{\it \small $^c$ The Niels Bohr Institute, Blegdamsvej 17, DK-2100 Copenhagen, Denmark}\\

\end{center}
\begin{abstract}
Starting from the leading soft term of the $5$-point amplitude, involving  a graviton and two Kerr black holes, 
that factorises into the product of the elastic amplitude without the graviton and the leading soft factor, we
compute the infrared divergent contribution to the imaginary part of the two-loop eikonal.  
Then, using analyticity  and crossing symmetry, 
we determine the radiative contribution to the real part of the two-loop eikonal and from it the radiative part of the deflection angle for spins aligned to the orbital angular momentum, the loss of angular momentum and the zero frequency limit  of the energy spectrum for any spin and for any spin orientation.  For spin one we find perfect agreement  with 
recent results obtained with the supersymmetric worldline formalism.  
\end{abstract}
\end{titlepage}


\section{Introduction}
\label{sec:intro}
The recent detection of gravitational waves (GW) by the LIGO and Virgo collaborations \cite{LIGOScientific:2016aoc,LIGOScientific:2016sjg,LIGOScientific:2017bnn,LIGOScientific:2017ycc,LIGOScientific:2017vwq,LIGOScientific:2018mvr,LIGOScientific:2020ibl} emitted by binary black holes or neutron stars mergers is shedding new light on the nature of strongly gravitating systems existing in the universe. The increasing sensitivity of the detectors demands for an effort to develop new theoretical tools capable of high-precision predictions of GW waveform templates. For these reasons, recently there has been considerable interest in the use of amplitude-based techniques for studying the inspiral phase of binary non-spinning massive objects coalescence, extracting the conservative part of the Effective-One-Body (EOB) potential~\cite{Buonanno:1998gg,Goldberger:2004jt,Goldberger:2016iau,Luna:2017dtq,Cheung:2018wkq,Kosower:2018adc,Bjerrum-Bohr:2018xdl,Bjerrum-Bohr:2019kec,Kalin:2019rwq,Kalin:2019inp,Cho:2021arx,Cristofoli:2020uzm,Mogull:2020sak,AccettulliHuber:2020dal}, classical observables~\cite{Kosower:2018adc,delaCruz:2020bbn,Cristofoli:2021vyo,Herrmann:2021tct,Cristofoli:2021jas}, waveforms and radiation~\cite{Damour:2017zjx,AccettulliHuber:2020dal,Mougiakakos:2021ckm,Jakobsen:2021smu,Mogull:2020sak,Herrmann:2021lqe,DiVecchia:2021bdo}.
Astrophysical objects are also characterized by their spin. Therefore, the post-Minkowskian (PM) expansion in the Newton's constant $G$  for studying the scattering of spinning objects has also received attention, using amplitudes methods and classical General Relativity (GR) analytical techniques~\cite{Arkani-Hamed:2017jhn,Guevara:2017csg,Bini:2017xzy,Vines:2017hyw,Bini:2018ywr,Vines:2018gqi,Guevara:2018wpp,Chung:2018kqs,Bautista:2019tdr,Maybee:2019jus,Guevara:2019fsj,Arkani-Hamed:2019ymq,Johansson:2019dnu,Chung:2019duq,Damgaard:2019lfh,Bautista:2019evw,Aoude:2020onz,Chung:2020rrz,Bern:2020buy,Aoude:2020ygw,Guevara:2020xjx,Liu:2021zxr,Kosmopoulos:2021zoq,Aoude:2021oqj,Jakobsen:2021lvp,Bautista:2021wfy,Chiodaroli:2021eug,Haddad:2021znf,Jakobsen:2021zvh,Saketh:2021sri,Chen:2021qkk,Jakobsen:2022fcj,Aoude:2022trd,Bern:2022kto,Adamo:2021rfq}. Furthermore, observables computed for scattering events in the PM expansion can be compared with those computed in the more traditional post-Newtonian (PN) \cite{Blanchet:2006zz} perturbation theory (for spin effects in the PN scheme see \textit{e.g.}~\cite{Levi:2014sba,Levi:2015msa,Levi:2015uxa,Levi:2015ixa,Levi:2016ofk,Levi:2019kgk,Levi:2020lfn,Levi:2020kvb,Levi:2020uwu,Kim:2021rfj}), which is more suitable for describing bound orbits, using the so-called Bound-to-Boundary (B2B) \cite{Kalin:2019rwq,Kalin:2019inp,Cho:2021arx} correspondence.

The contribution of the radiation reaction  in  the scattering of two spinless black holes  at 3PM has been crucial~\cite{DiVecchia:2020ymx} to get rid of    the divergence at high energy  that appears in the conservative part of the deflection angle~\cite{Bern:2019nnu,Bern:2019crd,Kalin:2020fhe}. This has been first shown in massive ${\cal{N}}=8$ supergravity by performing the explicit calculation
in the soft region~\cite{DiVecchia:2020ymx} instead of just the conservative region~\cite{Parra-Martinez:2020dzs}.  Immediately after  the radiation reaction was also computed in GR with two different methods. The first, discussed  in Ref.~\cite{Damour:2020tta}, is based  on the computation of   the loss of angular momentum that is then inserted   in a formula derived in Ref.~\cite{Bini:2012ji} that gives the radiative part of the deflection angle.  The second one~\cite{DiVecchia:2021ndb} uses instead the leading soft part  of the $5$-point amplitude for the emission of a graviton that
allows to compute the infrared divergent contribution to  the imaginary part of the eikonal at 3PM order. Then, using considerations based on analyticity and crossing symmetry,  the real part of the radiative eikonal    at 3PM  is  also extracted  and from it  the radiation reaction part of the deflection angle at 3PM has been obtained. The two approaches give  the same result that, at high energy, is also in agreement with the old result of Ref.~\cite{Amati:1990xe}.  This result is also  confirmed~\cite{Bjerrum-Bohr:2021vuf,Bjerrum-Bohr:2021din,Brandhuber:2021eyq} by extracting from the explicit two-loop calculations the  radial action  and from it the deflection angle.

In this paper we generalise the approach of Ref.~\cite{DiVecchia:2021ndb} to the  case of Kerr black holes  obtaining  a closed expression for the radiation reaction part of the deflection angle in the aligned-spin case and to all order in spin. Assuming that the logarithmic divergence appearing in the radiation reaction contribution to the scattering angle in the ultra-relativistic limit cancels with that appearing in its conservative part, we conjecture a formula describing the high-energy limit of the latter.  Furthermore, using the Bini-Damour linear response equation \cite{Bini:2012ji,Damour:2020tta,Bini:2021gat}, we find the angular momentum loss for any spin configuration and to all orders in spin. Our formula, for spin  one, agrees with the one computed in Ref.~\cite{Jakobsen:2021lvp,Jakobsen:2022fcj}.Furthermore, it is also consistent  with the general result for the angular momentum loss  obtained in Ref.~\cite{DiVecchia:2022owy}.

The paper is organized as follows. We start in section \ref{sec:tree } by introducing the kinematical setup describing the elastic $(2\rightarrow 2)$ and inelastic $(2\rightarrow 3)$ scattering processes under considerations and by showing the corresponding tree-level four-point and five-point amplitudes. We approximate the latter by using Weinberg's soft graviton theorem, that allows to factorize it in terms of the four-point amplitude and the leading soft factor. In section \ref{sec:soft} we introduce the main ideas of our computation and then we proceed to explicitly calculate the infrared divergent contribution  to  the imaginary part of two-loop eikonal $\delta_2$ using the three-particle unitarity cut. Then from it, using  analyticity and crossing symmetry, we derive the radiation reaction  piece of the real part of $\delta_2$. Our result shows that all the spin dependence is encoded in a single vector field, denoted by $\mathbf{f}$ in \eqref{F15}. We discuss how this feature can be understood in terms of a Newman-Janis shift \cite{Newman:1965tw}, which relates the Kerr to the Schwarzschild solution.
We end section \ref{sec:soft} by displaying the zero-frequency limit of the emitted energy spectrum. In section \ref{sec:rad} we determine the radiative contribution to the 3PM deflection angle for an aligned-spin configuration and the 2PM angular momentum loss in any spin configuration. Our final equations \eqref{F34}, \eqref{F28} and \eqref{F29} hold to all orders in spin.  We close in section \ref{sec:conc} with a short summary and an outlook. In this paper we use the mostly plus signature $\eta_{\mu\nu}=\mathrm{diag}(-1,1,1,1)$ and we mainly follow the conventions used in \cite{DiVecchia:2021ndb}.

\section{Tree-level four-point and soft five-point amplitudes}
\label{sec:tree }
\begin{figure}[h]
  \centering
  \begin{minipage}[b]{0.45\textwidth}
    \includegraphics[width=\textwidth]{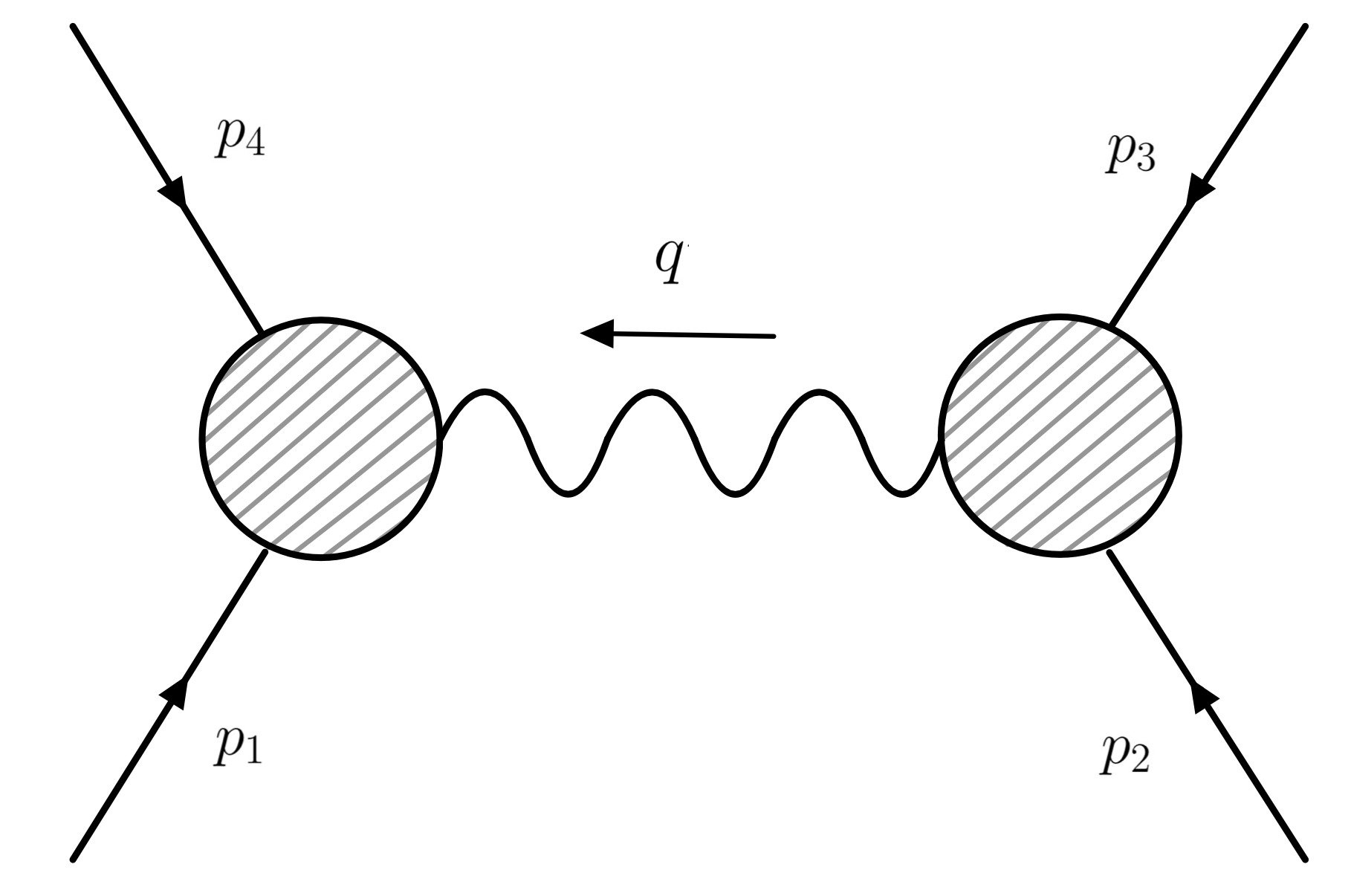}
  \end{minipage}
   \caption{t-channel for the one-graviton exchange.}
   \label{Fig1}
\end{figure}

We start by considering the tree-level $4$-point amplitude for elastic scattering of two Kerr black holes with masses $m_1$ and $m_2$ anxd rescaled spin vectors $a_1^{\mu}=S_1^{\mu}/m_1$ and $a_2^{\mu}=S_2^{\mu}/m_2$, that can be written in an exponentiated form as \cite{Guevara:2018wpp,Bautista:2019tdr,Guevara:2019fsj}
\begin{align}
\label{F1}
\mathcal{A}^{\mathrm{tree}}_4(q)=-\frac{\kappa^2m^2_1m^2_2}{t}\sigma^2\sum_{\pm}(1\pm v)^2\mathrm{exp}\bigg(\pm i\frac{\epsilon_{\mu\nu\alpha\beta}q^{\mu}a^{\nu}p_1^{\alpha}p_2^{\beta}}{m_1m_2\sigma v}\bigg)+\mathcal{O}(t^0),
\end{align}
where in the amplitude with one-graviton exchange we keep only terms non-analytic for $t=-q^2\sim0$ that are relevant 
for long-range effects. 
Here $\kappa=\sqrt{8\pi G}$ where $G$ is the Newton constant, $a=a_1+a_2$, $p_1$ and $p_2$ are the incoming momenta, $q=-p_1-p_4=p_2+p_3$ is the transferred momentum and $\sigma=\frac{1}{\sqrt{1-v^2}}=-\frac{p_1\cdot p_2}{m_1m_2}$. The amplitude in \eqref{F1}, which is the main building block of our analysis, can be obtained by gluing two massive spinning three-point amplitudes at minimal coupling \cite{Arkani-Hamed:2017jhn,Vines:2017hyw,Bautista:2019tdr}, as shown in Fig.\ref{Fig1}. 

Note that we treat all vectors as formally ingoing. We work in the center-of-mass frame, defined by
\begin{align}
\label{F2}
p_1=(E_1,\mathbf{p}),\qquad p_2=(E_2,-\mathbf{p}),\qquad\left|\mathbf{p}\right|=p,
\end{align}
where $E_i=\sqrt{m^2_i+p^2}$ and the total energy, $E=E_1+E_2= \sqrt{s}$ ($s$ being the Mandelstam variable $s=-(p_1+p_2)^2$), satisfies $Ep=m_1m_2\sqrt{\sigma^2-1}$. The spin vectors are given by
\begin{align}
\label{F3}
a_1=\bigg(\frac{\mathbf{p}\cdot\mathbf{a}_1}{m_1},\mathbf{a}_1+\frac{\mathbf{p}\cdot\mathbf{a}_1}{m_1(m_1+E_1)}\mathbf{p}\bigg),\quad a_2=\bigg(-\frac{\mathbf{p}\cdot\mathbf{a}_2}{m_2},\mathbf{a}_2+\frac{\mathbf{p}\cdot\mathbf{a}_2}{m_2(m_2+E_2)}\mathbf{p}\bigg),
\end{align}
so that $a_i^2=\mathbf{a}_i^2$. We choose the spatial component of the momenta and of the  impact parameter to be along the $x^3$ and $x^1$ directions, respectively, \textit{i.e.} $\mathbf{p}=p\,\mathbf{e}_3$ and $\mathbf{b}=b\,\mathbf{e}_1$, so that the asymptotic scattering plane is the $x^1x^3$ plane. Here $\mathbf{e}_i$ are the Cartesian spatial unit vectors. So far we have not specified the directions of the spatial spin vectors $\mathbf{a}_i$. If we choose them to be perpendicular to the $x^1x^3$ plane we are considering an aligned-spin configuration. However we will leave these directions arbitrary in the following, unless otherwise specified.

The exponential factor appearing in the amplitude \eqref{F1} in the center-of-mass frame can be expressed in terms of spatial vectors only as
\begin{align}
\label{F4}
\pm i\frac{\epsilon_{\mu\nu\alpha\beta}q^{\mu}a^{\nu}p_1^{\alpha}p_2^{\beta}}{m_1m_2\sigma v}=\pm i\mathbf{q}\cdot(\hat{\mathbf{p}}\times\mathbf{a}),\qquad \hat{\mathbf{p}}=\frac{\mathbf{p}}{p},
\end{align}
where $\mathbf{q}$ is the $2$-dimensional spatial transferred momentum satisfying $\mathbf{q}\cdot\mathbf{p}=0$ ($\epsilon_{0123}=1$). We now consider the inelastic $(2\rightarrow 3)$ process with the emission of an additional graviton of momentum $k$ and we denote the corresponding $5$-point amplitude by $\mathcal{A}_5^{\mu\nu}(q,k)$. We use the following parametrisation for the momenta 
\begin{align}
\label{F5}
&p_1\simeq\bar{p}_1-\frac{1}{2}q,\quad p_2\simeq\bar{p}_2+\frac{1}{2}q,\quad p_3\simeq-\bar{p}_2+\frac{1}{2}q,\quad p_4\simeq-\bar{p}_1-\frac{1}{2}q,
\end{align}
where $p_1^2=m^2_1\simeq\bar{p}^2_1 $, $p_2^2=m_2^2\simeq\bar{p}^2_2$ and $\bar{p}_1\cdot q\simeq 0$, $\bar{p}_2\cdot q\simeq 0$ hold up to terms proportional to $k$, which are negligible in the soft $k\rightarrow 0$ limit. In this regime, the five point amplitude for such process factorizes according to Weinberg soft graviton theorem \cite{Weinberg:1965nx} as
\begin{align}
\label{F6}
\mathcal{A}_5^{\mu\nu}(q,k)=\kappa\sum_{i=1}^4\frac{p_i^{\mu}p_i^{\nu}}{k\cdot p_i}\mathcal{A}^{\mathrm{tree}}_4(q)+\mathcal{O}(k^0).
\end{align}
The multiplicative factor is the so-called Weinberg pole, that diverges as $k^{-1}$ for small $k$. In general, one could go further in the soft expansion and consider subleading soft terms with higher powers of $k$ \cite{Cachazo:2014fwa,Bern:2014vva}. However, since we shall only be concerned in the infrared divergent part of the imaginary part of the two-loop eikonal $\delta_2$, which is responsible for radiation reaction effects \cite{DiVecchia:2021ndb,DiVecchia:2021bdo}, we can neglect all such terms and consider contributions that come only from those diagrams where the graviton line is attached to an external leg. 

By taking the classical limit, corresponding to $q\rightarrow 0$, and therefore expanding the soft factor in powers of $q$ and keeping only the terms linear in $q$, equation \eqref{F6} becomes
\begin{align}
\label{F7}
\mathcal{A}_5^{\mu\nu}(q,k)=\kappa\bigg(\frac{\bar{p}_1^{\mu}\bar{p}_1^{\nu}}{(\bar{p}_1\cdot k)^2}(q\cdot k)-\frac{(\bar{p}_1^{\mu}q^{\nu}+\bar{p}_1^{\nu}q^{\mu})}{\bar{p}_1\cdot k}
-1\leftrightarrow 2\bigg)\mathcal{A}_4^{\mathrm{tree}}(q)+\mathcal{O}(q^2,k^0).
\end{align}
where we have dropped higher powers of $q$ that are negligible in the classical limit.

\section{Infrared divergence of the two-loop eikonal}
\label{sec:soft}

In this section we first go to impact parameter space and then we compute the  infrared divergent  contribution to the unitarity relation given by  the three-particle cut that provides  the imaginary part of the two-loop eikonal. Then following 
the analysis based on crossing symmetry and analyticity  discussed in Refs.~\cite{DiVecchia:2020ymx,DiVecchia:2021ndb,DiVecchia:2021bdo} we can extract  the radiation reaction contribution to the real part of the eikonal  by means of the relation
\begin{align}
\label{F8}
\mathrm{Re}\,2\delta_2^{\mathrm{rr}}=-\lim_{\epsilon\to 0}[\pi\epsilon(\mathrm{Im}\,2\delta_2)],
\end{align}
where $\epsilon=\frac{4-D}{2}$. Equation \eqref{F8} has been confirmed by explicit two-loop calculations in massive ${\cal{N}}=8$ supergravity and  in GR \cite{DiVecchia:2021bdo} and is believed to hold at all energies and for all gravitational theories involving two derivatives. As a consequence of \eqref{F8}, the infrared divergent part of the tree-level, on-shell, inelastic amplitude for the $(2\rightarrow 3)$ process under consideration, which is simply fixed  in \eqref{F7} by the leading Weinberg  soft graviton theorem, is the only necessary ingredient to study radiation reactions effects \cite{DiVecchia:2021ndb,Heissenberg:2021tzo}. This method provides a powerful shortcut to get information about $\mathcal{O}(G^3)$ classical  observables, which are usually obtained by a more involved two-loop computation. In particular, the two observables that one can extract from Eq. \eqref{F8} are the zero-frequency limit of the emitted energy spectrum \cite{Smarr:1977fy} and the radiative contribution to the 3PM scattering angle given by 
\begin{align}
\label{F8.5}
\left.\frac{d E^{\mathrm{rad}}}{d\omega}\right|_{\omega\to 0}=\lim_{\epsilon\to 0}[-4\hbar\epsilon (\mathrm{Im}\,2\delta_2)],\qquad \theta^{\mathrm{rr}}_{3}=-\frac{\hbar}{p}\frac{\partial \mathrm{Re}\,2\delta_2^{\mathrm{rr}}}{\partial b}.
\end{align}
Furthermore, having determined  $\theta_3^{\mathrm{rr}}$  from Eq. \eqref{F8.5} and assuming the 
Bini-Damour linear response equation~\cite{Bini:2012ji,Damour:2020tta,Bini:2021gat,Veneziano:2022zwh,Manohar:2022dea} that gives it in terms of the angular momentum $J^{\mathrm{loss}}$ and the conservative deflection angle $\theta^{\mathrm{cons}}$, 
\begin{align}
\label{F8.6}
\theta^{\mathrm{rr}}=-\frac{1}{2}\frac{\partial \theta^{\mathrm{cons}}}{\partial E}E^{\mathrm{loss}}-\frac{1}{2}\frac{\partial \theta^{\mathrm{cons}}}{\partial J}J^{\mathrm{loss}},
\end{align}
where  $J=bp$ is the initial angular momentum in the center-of-mass frame, one can extract, inverting \eqref{F8.6} and discarding the contribution of the energy loss that starts at $\mathcal{O}(G^3)$, the contribution to the angular momentum loss at $\mathcal{O}(G^2)$, denoted here by $J^{\mathrm{loss}}_2$, which is entirely given by zero-frequency gravitons \cite{DiVecchia:2022owy}. 

Note that the scattering angle can be only defined in the aligned-spin configuration.
Indeed, when the two spin vectors are not aligned, the scattering dynamics is not planar and the asymptotic scattering planes at $t=-\infty$ and $t=\infty$ will not coincide and there are rather two scattering angles \cite{Bern:2020buy}. In this case, the angular momentum in the center-of-mass frame is radiated in both directions orthogonal to $\mathbf{p}$ \cite{Jakobsen:2021zvh}. In our kinematics setup described after equations \eqref{F3}, it means that the radiated angular momentum has non-vanishing components along $x^1$ and $x^2$. In the rest of this section we compute the infrared divergent part of $\mathrm{Im}\,2\delta_2$.

We first take the leading soft $5$-point amplitude in Eq. \eqref{F7} and we go to impact parameter space by the following formula:
\begin{align}
\label{F9}
\tilde{\mathcal{A}}_5^{\mu\nu}(b,k)=\int\frac{d^{4-2\epsilon}q}{(2\pi)^{4-2\epsilon}}e^{iq\cdot b}\hat{\delta}(2q\cdot \bar{p}_1)\hat{\delta}(2q\cdot \bar{p}_2)\mathcal{A}^{\mu\nu}_5(q,k),
\end{align}
where $\hat{\delta}(2q\cdot \bar{p}_i)\equiv 2\pi\delta(2q\cdot \bar{p}_i)$ enforces the orthogonality conditions $q\cdot \bar{p}_i=0$. Because of the two $\delta$-functions the integral over the two longitudinal components spanned by  $\bar{p}_i$ can be easily performed and one arrives at
\begin{align}
\label{F10}
\tilde{\mathcal{A}}^{\mu\nu}_5(b,k)=\frac{1}{4m_1m_2\sqrt{\sigma^2-1}}\int\frac{d^{2-2\epsilon}\mathbf{q}}{(2\pi)^{2-2\epsilon}}e^{i\mathbf{q}\cdot \mathbf{b}}\mathcal{A}^{\mu\nu}_5(\mathbf{q},k).
\end{align}
Inserting the amplitude \eqref{F7} in the above integral and using the representation in \eqref{F4} for the spin exponential factor in the center-of-mass frame we see that the entire spin dependence of $\tilde{\mathcal{A}}^{\mu\nu}(\mathbf{b},k)$ is encoded in the shift $\mathbf{b}\rightarrow\mathbf{b}\pm \hat{\mathbf{p}}\times\mathbf{a}$. This property had been already noticed in \cite{Arkani-Hamed:2019ymq,Guevara:2020xjx,Monteiro:2021ztt}\footnote{See also \cite{Monteiro:2014cda,Luna:2016due} for earlier connected works.} and it comes from the nature of minimally coupled amplitudes introduced in \cite{Arkani-Hamed:2017jhn} and from the Fourier factor $e^{i\mathbf{q}\cdot \mathbf{b}}$ relating classical observables in impact parameter space to amplitudes \cite{Kosower:2018adc,Maybee:2019jus}. Ultimately, it admits an interpretation in terms of a Newman-Janis shift \cite{Newman:1965tw}, relating the spacetime of a spinning black hole to that of a spinless one.

\begin{figure}[h]
  \centering
  \begin{minipage}[b]{0.5\textwidth}
    \includegraphics[width=\textwidth]{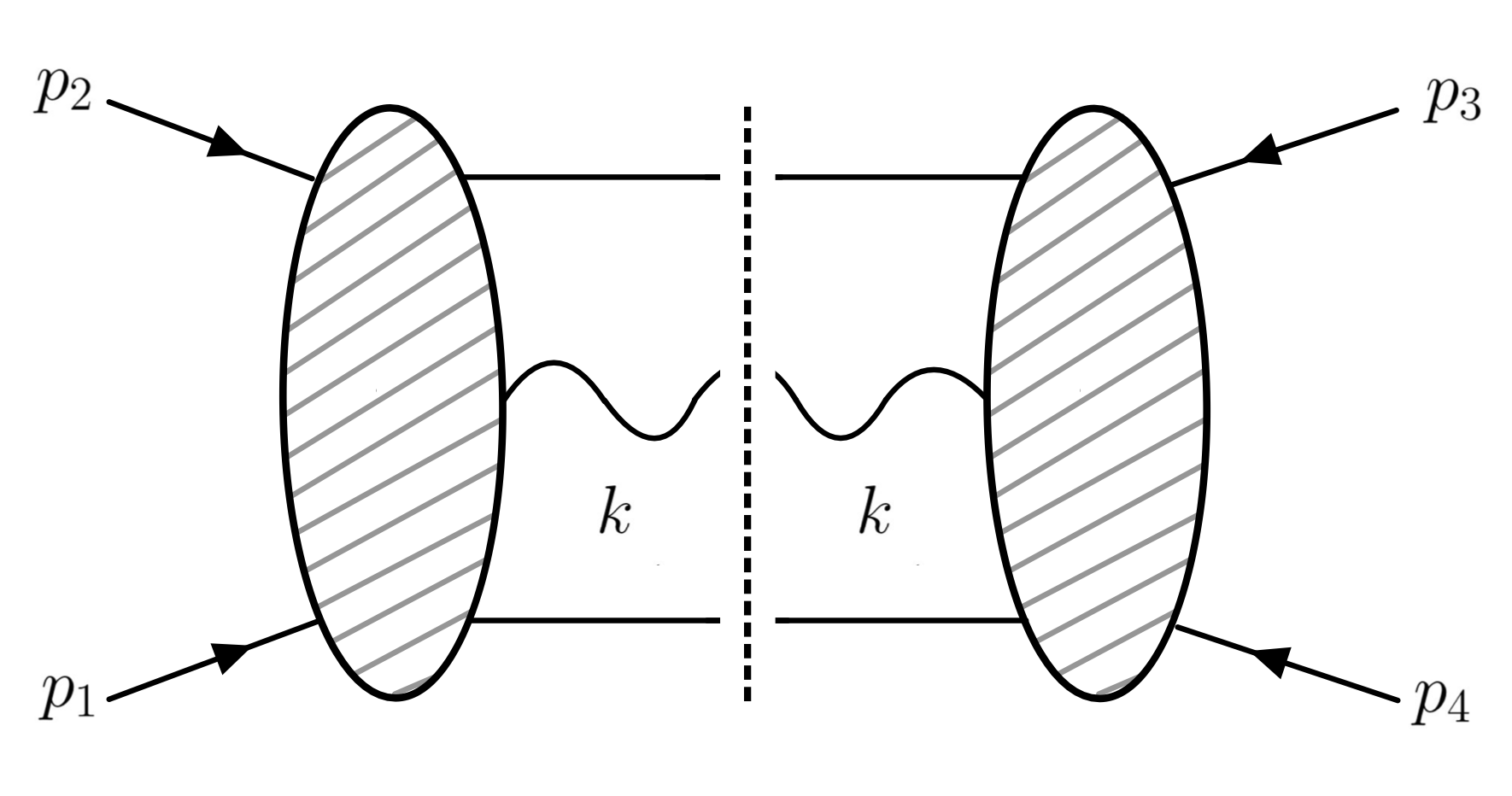}
  \end{minipage}
   \caption{Three-particle cut yielding the infrared divergent part of $\mathrm{Im}\,2\delta_2$ in \eqref{F13}}
   \label{Fig2}
\end{figure}

This allows to easily perform the above integral for all spins, yielding the following amplitude in impact parameter space
\begin{align}
\label{F11}
\tilde{\mathcal{A}}_5^{\mu\nu}(b,k)\simeq\frac{i\kappa^3}{8\pi}\frac{m_1m_2\sigma^2}{\sqrt{\sigma^2-1}}\sum_{\pm}\frac{(1\pm v)^2}{\mathbf{b}_{\pm}^2}\bigg(\frac{\bar{p}_1^{\mu}\bar{p}_1^{\nu}}{(\bar{p}_1\cdot k)^2}(\mathbf{k}\cdot \mathbf{b}_{\pm})-\frac{\bar{p}_1^{\mu}b_{\pm}^{\nu}+\bar{p}_1^{\nu}b_{\pm}^{\mu}}{\bar{p}_1\cdot k}-1\leftrightarrow 2\bigg),
\end{align}
where we defined for convenience $b^{\mu}_{\pm}\equiv (0,\mathbf{b}\pm \hat{\mathbf{p}}\times\mathbf{a})$. Note that this result is valid to all orders in spin and in any spin configuration. The multipole expansion in the spin vector is recovered from the expansion of $\mathbf{b}_{\pm}^{-2}$ around $\mathbf{a}=0$ as
\begin{align}
\label{F12}
\frac{1}{\mathbf{b}_{\pm}^2}=\frac{1}{b^2}\mp\frac{\mathbf{b}\cdot (\hat{\mathbf{p}}\times\mathbf{a})}{b^4}-\frac{(\hat{\mathbf{p}}\times\mathbf{a})^2}{b^4}+2\frac{[\mathbf{b}\cdot(\hat{\mathbf{p}}\times\mathbf{a})]^2}{b^6}+\mathcal{O}(\mathbf{a}^3).
\end{align}
We are now ready to compute the infrared divergent part of $\mathrm{Im}\,2\delta_2$ using the three-particle unitarity cut \cite{Amati:2007ak,DiVecchia:2020ymx,DiVecchia:2021ndb} depicted in Fig.\ref{Fig2} as 
\begin{align}
\label{F13}
\mathrm{Im}\,2\delta (b,\sigma) = \frac{1}{2} \int \frac{d^{3-2\epsilon} \mathbf{k}}{2\left|\mathbf{k}\right|(2\pi)^{3-2\epsilon} }  \tilde{\mathcal{A}}_5^{\mu\nu}(b,k)P_{\mu\nu;\rho\sigma}\tilde{\mathcal{A}}_5^{*\rho\sigma}(b,k),
\end{align} 
 where $\omega=\left|\mathbf{k}\right|$ and
 we use dimensional regularisation to
capture the infrared divergence in 
$\mathrm{Im}\,2\delta_2$.

 The de Donder projector $P_{\mu\nu;\rho\sigma}=\frac{1}{2}\eta_{\mu\rho}\eta_{\nu\sigma}+\frac{1}{2}\eta_{\mu\sigma}\eta_{\nu\rho}-\frac{1}{2}\eta_{\mu\nu}\eta_{\rho\sigma}$ has been inserted in \eqref{F13} in order to single out only the physical, transverse-traceless projection of $\tilde{\mathcal{A}}_5^{\mu\nu}$.

We parametrize the momentum of the emitted graviton\footnote{In our conventions $k$ is ingoing.} as 
$k=-\omega(1,\hat{\mathbf{x}})$ where $\hat{\mathbf{x}}=(\sin\theta,\sin\theta\sin\varphi,\cos\theta)$ and hence $\left|\mathbf{k}\right|=\omega$. The various scalar product are
 \begin{align}
\label{F13a}
&\mathbf{k}\cdot \mathbf{b}_{\pm} =-\omega\sin\theta[b\cos\varphi\pm (a^1\sin\varphi-a^2\cos\varphi)],\\&\bar{p}_1\cdot k=\omega(E_1-p\cos\theta),\qquad\bar{p}_2\cdot k=\omega(E_2+p\cos\theta),\\
&\mathbf{b}_{\pm}^2=b^2+(a^1)^2+(a^2)^2\mp 2ba^2.
\end{align}
where $\mathbf{a}=a^i\mathbf{e}_i$. Note that the result will not depend on $a^3$, component of $\mathbf{a}$ along $\mathbf{p}$. Integrating over $\omega$ and over $\varphi$ and keeping only the term proportional to $\epsilon^{-1}$  we get for the infrared divergent part of the imaginary part of the two-loop eikonal\footnote{In equation \eqref{F14} we restored $\hbar$, that was set equal to one so far.}
\begin{align}
\label{F14}
\mathrm{Im}\,2\delta_2(\sigma,b)&\simeq-\frac{1}{2\epsilon}\frac{\pi}{2\hbar(2\pi)^3}\bigg(\frac{\kappa^3m_1m_2\sigma^2}{8\pi\sqrt{\sigma^2-1}}\bigg)^2\bigg(\sum_{\pm}\frac{(1\pm v)^2\mathbf{b}_{\pm}}{\mathbf{b}^2_{\pm}}\bigg)^2\mathcal{I}(\sigma),
\end{align}
where \cite{Damour:2020tta}\footnote{Our definition of $\mathcal{I}(\sigma)$ has an additional factor $2$ with respect to the one in \cite{Jakobsen:2021lvp}.}
\begin{align}
\nonumber\mathcal{I}(\sigma)=&\int_{-1}^1\text{d}x\bigg\{\frac{1-x^2}{4}\bigg[\frac{m_1^4}{(E_1-\bar{p}x)^4}+\frac{m_2^4}{(E_2+\bar{p}x)^4}-\frac{2m_1^2m_2^2(2\sigma^2-1)}{(E_1-\bar{p}x)^2(E_2+\bar{p}x)^2}\bigg]\\\nonumber &-2\bigg[\frac{m_1^2}{(E_1-\bar{p}x)^2}+\frac{m_2^2}{(E_2+\bar{p}x)^2}-\frac{2m_1m_2\sigma}{(E_1-\bar{p}x)(E_2+\bar{p}x)}\bigg]\bigg\}\\&=\frac{2}{\sigma^2-1}\bigg[\frac{8-5\sigma^2}{3}-\frac{\sigma(3-2\sigma^2)}{\sqrt{\sigma^2-1}}\cosh^{-1}(\sigma)\bigg].
\end{align}
Introducing the notation $\beta(\sigma)=2m_1m_2G(2\sigma^2-1)$ as in \cite{DiVecchia:2021ndb,DiVecchia:2021bdo} and the spatial vector $\mathbf{f}$ as
\begin{align}
\label{F15}
\sum_{\pm}\frac{(1\pm v)^2\mathbf{b}_{\pm}}{\mathbf{b}_{\pm}^2}\equiv \frac{2(2\sigma^2-1)}{\sigma^2b}\mathbf{f}(a,b,\sigma),
\end{align}
the final result for the infrared divergent part of $\mathrm{Im}\,2\delta_2$ can be simply written as
\begin{align}
\label{F16}
\mathrm{Im}\,2\delta_2(\sigma,b)\simeq-\frac{1}{2\epsilon}\frac{G\beta^2(\sigma)}{2\pi\hbar(\sigma^2-1) b^2}\mathcal{I}(\sigma)\mathbf{f}^2(a,b,\sigma).
\end{align}
When the spins are parallel to the orbital angular momentum the vector $\mathbf{f}$ is 
\begin{align}
\label{F23}
\mathbf{f}(a,b,\sigma)\equiv f(a,b,\sigma)\frac{\mathbf{b}}{b},\qquad f(a,b,\sigma)=\frac{1+\frac{2\sigma\sqrt{\sigma^2-1}}{2\sigma^2-1}\frac{a}{b}}{1-\left(\frac{a}{b}\right)^2}.
\end{align}
For arbitrary spin orientation, $\mathbf{f}$ has non-vanishing components also along $\hat{\mathbf{p}}\times\mathbf{a}$ and explicit expressions for $\mathbf{f}$ and $\mathbf{f}^2$ can be found in Appendix \ref{AA}. 

Using Eq. \eqref{F8} we can extract the radiative contribution to the real part of the two-loop eikonal   
\begin{align}
\label{F17}
\mathrm{Re}\,2\delta_2^{\mathrm{rr}}(\sigma,b)=\frac{G\beta^2(\sigma)}{4\hbar(\sigma^2-1)b^2}\mathcal{I}(\sigma)\mathbf{f}^2(a,b,\sigma)=\mathrm{Re}\,2\delta_2^{\mathrm{rr}}(\sigma,b)\big|_{\mathbf{a}=0}\mathbf{f}^2(a,b,\sigma),
\end{align}
This equation 
shows that the radiation reaction part of the eikonal of two Kerr black-holes scattering is simply obtained from that of two Schwarzschild black holes by means of a Newman-Janis shift, as stressed after \eqref{F10}. In particular, we find a simple rule that allows to get the result for spinning black holes from that of spinless ones. We start to notice that, for the latter,
the combination $[2(2\sigma^2+1)\mathbf{b}/\sigma^2b^2]^2$ that appears in $\mathrm{Re}\,2\delta_2$ can be rewritten as
\begin{align}
\label{F18}
\left(\frac{2(2\sigma^2-1)}{\sigma^2 b^2}\mathbf{b}\right)^2=\bigg(\frac{(1+v)^2\mathbf{b}}{\mathbf{b}^2}+\frac{(1-v)^2\bar{\mathbf{b}}}{\bar{\mathbf{b}}^2}\bigg)^2,
\end{align}
where we introduced an auxiliary vector $\bar{\mathbf{b}}$ such that $\bar{\mathbf{b}}\stackrel{\mathbf{a}=0}{=}\mathbf{b}$. Then to get the result for $\mathbf{a}\neq 0$ we just need to do the replacements
\begin{align}
\label{F19}
\mathbf{b}\stackrel{\mathbf{a}\neq 0}\longrightarrow \mathbf{b}_+,\qquad\bar{\mathbf{b}}\stackrel{\mathbf{a}\neq 0}\longrightarrow \mathbf{b}_-.
\end{align}
We complete this section by using the first equation of \eqref{F8.5}  for computing 
the zero frequency limit of the energy spectrum 
\begin{align}
\label{F20}
\frac{d E^{\mathrm{rad}}}{d \omega}\bigg|_{\omega\to 0}=\frac{4G^3m_1^2m_2^2(2\sigma^2-1)^2}{\pi b^2(\sigma^2-1)}\mathcal{I}(\sigma)\mathbf{f}^2(a,b,\sigma),
\end{align}
that can be expanded in the static (PN) limit $\sigma\rightarrow 1$ yielding,
\begin{align}
\label{F21}
\frac{d E^{\mathrm{rad}}}{d \omega}\bigg|_{\omega\to 0}\simeq\frac{32G^3m_1^2m_2^2}{5\pi}\frac{b^2}{(\mathbf{b}+\hat{\mathbf{p}}\times\mathbf{a})^2(\mathbf{b}-\hat{\mathbf{p}}\times\mathbf{a})^2},
\end{align}
that gives the all orders in spin correction to the original Smarr formula \cite{Ruffini:1970sp,Smarr:1977fy}. Further expanding up to quadratic order in spin we find
\begin{align}
\label{F22}
\frac{d E^{\mathrm{rad}}}{d \omega}\bigg|_{\omega\to 0}\simeq\frac{32G^3m_1^2m_2^2}{5\pi b^2}\left(1+2\frac{(a^2)^2-(a^1)^2}{b^2}\right).
\end{align}
When the spins are orthogonal to the scattering plane, Eq. \eqref{F21} reduces to 
\begin{align}
\label{F22.A}
\frac{d E^{\mathrm{rad}}}{d \omega}\bigg|_{\omega\to 0}\simeq\frac{32G^3m_1^2m_2^2}{5\pi b^2}\frac{1}{\left[1-\left(\frac{a}{b}\right)^2\right]^2}.
\end{align}

\section{Radiative scattering angle and  angular momentum loss to all orders in spin}
\label{sec:rad}

In this section we start studying the  case in which the two spins are aligned  \textit{i.e.} we set $a^1=0$ and we define $a\equiv a^2$ to simplify the notation and from the real part of the eikonal in \eqref{F17} we extract the radiative part of the scattering angle  to all orders in spin.
In this case the vector $\mathbf{f}$ is given in \eqref{F23} so that the real part of the radiation reaction eikonal becomes
\begin{align}
\label{F24}
\mathrm{Re}\,2\delta_2^{\mathrm{rr}}(\sigma,b)=\frac{G\beta^2(\sigma)}{4\hbar(\sigma^2-1)b^2}\mathcal{I}(\sigma)f^2(a,b,\sigma),
\end{align}
and $\theta^{\mathrm{rr}}_3$, using \eqref{F8.5} is
\begin{align}
\label{F25}
\theta^{\mathrm{rr}}_3(\sigma,b)=\frac{G\beta^2(\sigma)}{2(\sigma^2-1)pb^3}\mathcal{I}(\sigma)\frac{(1+\frac{2\sigma\sqrt{\sigma^2-1}}{2\sigma^2-1}\frac{a}{b})\left[1+\frac{4\sigma\sqrt{\sigma^2-1}}{2\sigma^2-1}\frac{a}{b}+\left(\frac{a}{b}\right)^2\right]}{\left[1-\left(\frac{a}{b}\right)^2\right]^3}.
\end{align}
In the case of spin $1$ that contains only terms up to the order $a^2$ the previous expression is equal to
\begin{equation}
\theta^{\mathrm{rr}}_3(\sigma,b) =\frac{G\beta^2(\sigma)}{2(\sigma^2-1)pb^3}\mathcal{I}(\sigma) \Bigg( 1 + \frac{6 \sigma \sqrt{\sigma^2-1}}{2\sigma^2-1} \frac{a}{b} + 4 \frac{ 6 \sigma^4 -6 \sigma^2+1}{(2\sigma^2-1)^2}
\frac{a^2}{b^2}\Bigg),
\label{F26a}
\end{equation}
 that agrees with Eq. (19) of \cite{Jakobsen:2022fcj} for $s_{E,+}=0$. For spin $1$ the authors of Ref.~\cite{Jakobsen:2022fcj} checked that the deflection angle is not divergent at high energy because the logarithmic divergence that one finds in the conservative part is cancelled by the logarithmic divergence appearing in the radiative part in analogy with what happens for spin zero. Assuming that the same cancellation also happens for arbitrary spin, from  the high energy behaviour of the real part of the two-loop eikonal
given by 
\begin{align}
\label{F32}
\mathrm{Re}\,2\delta_2^{\mathrm{rr}}(\sigma,b)\sim \frac{16G^3m_1^2m_2^2}{(1-\frac{a}{b})^2}\sigma^2\log\sigma,\qquad\sigma\to\infty,
\end{align}
 we can  deduce the high energy limit of the conservative part of the real part of the two-loop eikonal
 \begin{align}
\label{F33}
\mathrm{Re}\,2\delta_2^{\mathrm{cons}}(\sigma,b)\sim -\frac{16G^3m_1^2m_2^2}{(1-\frac{a}{b})^2}\sigma^2\log\sigma,\qquad\sigma\to\infty.
\end{align} 
This implies that the high energy limit of the conservative scattering angle must be equal to
\begin{align}
\label{F34}
\theta^{\mathrm{cons}}_3(\sigma,b)\simeq -\frac{32 G^3Em_1m_2}{(1-\frac{a}{b})^3}\sigma\log\sigma,\qquad\sigma\to\infty,
\end{align}
The previous equations are valid for any spin and in particular Eq. \eqref{F34} agrees with the high energy limit of Eqs.
(16a), (16b) and (16c) of Ref.~\cite{Jakobsen:2022fcj}.

Then we want to use Eq. \eqref{F8.6} to determine the  2PM loss of  angular momentum from the radiative scattering angle. To do so we need the 1PM conservative scattering angle  
$\theta_1^{\mathrm{cons}}$~\cite{Guevara:2018wpp,Guevara:2019fsj}  
\begin{align}
\label{F27}
\theta_1^{\mathrm{cons}}(\sigma,b)=\frac{\beta(\sigma)}{pb\sqrt{\sigma^2-1}}f(a,b,\sigma)=\frac{\beta(\sigma)}{J\sqrt{\sigma^2-1}}\left(\frac{1+\frac{2\sqrt{\sigma^2-1}}{2\sigma^2-1}\frac{p a}{J}}{1-\left(\frac{pa}{J}\right)^2}\right).
\end{align}
that  is given by  $-\frac{1}{p}\frac{\partial\chi_1}{\partial b}$, where $\chi_1$ is the 1PM eikonal, obtained by inserting  the tree-level $4$-point amplitude  of  \eqref{F1} in Eq. \eqref{F10}.

We find that the 2PM loss of  angular momentum in the center-of-mass frame is equal to 
\begin{align}
\label{F28}
\mathbf{J}^{\mathrm{loss}}_2(\sigma,b)=J\frac{G\beta(\sigma)}{b^2\sqrt{\sigma^2-1}}\mathcal{I}(\sigma)\hat{\mathbf{p}}\times \left(f(a,b,\sigma)\frac{\mathbf{b}}{b}\right)=J\frac{G\beta(\sigma)}{b^2\sqrt{\sigma^2-1}}\mathcal{I}(\sigma)f(a,b,\sigma)\mathbf{e}_2.
\end{align}
Note that the only non-vanishing component of the 2PM angular momentum loss is  along the $\mathbf{e}_2$ direction, perpendicular to the scattering plane. This is due to the fact that in the aligned-spin case the scattering dynamics is planar, just as in the spinless scenario.

We note that \eqref{F28} admits a natural generalisation in the case of non-aligned spin as\footnote{We have also confirmed equation \eqref{F29} with an explicit computation of the angular momentum loss using the results in \cite{DiVecchia:2022owy}.}
\begin{align}
\label{F29}
\mathbf{J}^{\mathrm{loss}}_2(\sigma,b)=J\frac{G\beta(\sigma)}{b^2\sqrt{\sigma^2-1}}\mathcal{I}(\sigma)\hat{\mathbf{p}}\times \mathbf{f}(a,b,\sigma).
\end{align}
Contrarily to \eqref{F23}, in this case $\mathbf{f}$ does not lie entirely along $\mathbf{b}$, but has also one non-vanishing component along $\hat{\mathbf{p}}\times\mathbf{a}$, see Appendix \ref{AA}. Therefore, as already mentioned, the 2PM angular momentum loss in the center-of-mass frame  has components in both directions orthogonal to $\hat{\mathbf{p}}$. Even if \eqref{F29} has been obtained by generalising \eqref{F28} straightforwardly, it admits an interpretation as follows. Introducing the 1PM classical momentum transfer as $Q^{\mu}_1=-\frac{\partial \chi_1}{\partial b_{\mu}}$, we find $Q^0_1=0$ and{\footnote{We thank Carlo Heissenberg and Justin Vines for  discussions about this part of the paper.}
\begin{align}
\label{F30.2}
\mathbf{Q}_1(\sigma,b)=\frac{\beta(\sigma)}{b\sqrt{\sigma^2-1}}\mathbf{f}(a,b,\sigma),\qquad \hat{\mathbf{p}}\cdot \mathbf{Q}_1=0,
\end{align}
so that  \eqref{F29} can be more conveniently rewritten as
\begin{align}
\label{F30.3}
\mathbf{J}_2^{\mathrm{loss}}(\sigma,b)=G\,\mathcal{I}(\sigma)\,\mathbf{p}\times\mathbf{Q}_1(\sigma,b)=\mathbf{L}_2^{\mathrm{loss}}(\sigma,b)+\mathbf{S}_2^{\mathrm{loss}}(\sigma,b),
\end{align}
where
\begin{align}
\label{F30.4}
\mathbf{L}^{\mathrm{loss}}_2=G\,\mathcal{I}(\sigma)c_o(a,b,\sigma)\mathbf{p}\times\mathbf{b}\qquad\mathbf{S}^{\mathrm{loss}}_2(a,b,\sigma)=G\,\mathcal{I}(\sigma)c_s(a,b,\sigma)\mathbf{a},
\end{align}
where $c_o$ and $c_s$ have been defined in Appendix \ref{AA}.  The 2PM  angular momentum loss receives contributions from the variation of directions of the momenta and of the spin vectors of the particles involved in the scattering process. \\
Expanding equations \eqref{F28} and \eqref{F29} up to quadratic order in the spin we find respectively
\begin{align}
\label{F30}
\mathbf{J}^{\mathrm{loss}}_2(\sigma,b)=J\frac{G\beta(\sigma)}{b^2\sqrt{\sigma^2-1}}\mathcal{I}(\sigma)\left[1+\frac{2\sigma\sqrt{\sigma^2-1}}{2\sigma^2-1}\frac{a}{b}+\left(\frac{a}{b}\right)^2\right]\mathbf{e}_2,
\end{align}
for aligned spins and 
\begin{align}
\label{F31}
\nonumber\mathbf{J}^{\mathrm{loss}}_2(\sigma,b)&=J\frac{G\beta(\sigma)}{b^2\sqrt{\sigma^2-1}}\mathcal{I}(\sigma)\left[\bigg(-\frac{2\sigma\sqrt{\sigma^2-1}}{2\sigma^2-1}\frac{a^1}{b}-2\frac{a^1a^2}{b^2}\bigg)\mathbf{e}_1\right.\\&+\left.\bigg(1+\frac{2\sigma\sqrt{\sigma^2-1}}{2\sigma^2-1}\frac{a^2}{b}+\frac{(a^2)^2-(a^1)^2}{b^2}\bigg)\mathbf{e}_2,\right],
\end{align}
for non-aligned spins.
These equations exactly match equation (18) of \cite{Jakobsen:2022fcj} and equation (29) of \cite{Jakobsen:2021lvp} obtained with supersymmetric worldline formalism. Note that equation \eqref{F25}, \eqref{F28} and \eqref{F29} extend these results to all orders in spin. 

{\section{Conclusions and outlook}
\label{sec:conc}

From the leading soft term of a five-point amplitude involving two particles with spin minimally coupled to gravity that scatter producing also a low-energy graviton, we have computed the  infrared divergent piece of the imaginary part of the two-loop eikonal, $\delta_2$, generalising the procedure followed for spin zero in Ref.~\cite{DiVecchia:2021ndb}. Using then analyticity and crossing symmetry~\cite{DiVecchia:2021ndb} we have derived the radiation reaction contribution to the real part of the $\delta_2$ and from it the radiation reaction contribution to the deflection angle at 3PM ($\mathcal{O}(G^3)$). 

Using the Bini-Damour  relation in Eq. \eqref{F8.6}, we have determined the  angular momentum
loss  at $\mathcal{O}(G^2)$ for arbitrary spin configurations and to all orders in spin. 

Our results agree for spin zero with those in Refs.~\cite{Damour:2020tta,DiVecchia:2021ndb} and for spin one  with the complete calculation performed in Ref.~\cite{Jakobsen:2022fcj}}.

An interesting aspect of our results is that spin effects for soft bremsstrahlung are entirely encoded in the vector $\mathbf{f}$ in \eqref{F15}, that comes from a Newman-Janis shift in impact parameter space, as explained in \eqref{F19}. The origin of this  unexpected feature is that the entire calculation we have performed has the tree-level four-point amplitude in \eqref{F1} as starting point. In the latter, the spin structure appears as compactly organised in a single exponential factor, because of the nature of minimally coupled amplitudes. It would be interesting to understand to what extent such a simple criterion could be used to gather additional information about the conservative part of $\delta_2$, whose ultra-relativistic behaviour appears to be already fixed by the present analysis.

In equation \eqref{F20} we have shown the zero frequency limit of the emitted energy spectrum at $\mathcal{O}(G^3)$. The reason why we could do this is that we have only used the Weinberg's soft graviton theorem to determine the leading term of the inelastic five-point amplitude, because it was the necessary ingredient to fix the infrared divergence in $\mathrm{Im}\,2\delta_2$. In principle, one could go further in the soft expansion \cite{Cachazo:2014fwa,Bern:2014vva}, therefore considering also terms containing the total angular momentum operator, and understand how they contribute to the emitted energy and angular momentum spectrum beyond the leading soft limit.

\subsection*{Acknowledgements} 

We thank Luca Buoninfante, Alessandro Georgoudis, Kays Haddad, Carlo Heissenberg, Henrik Johansson, Alexander Ochirov, Paolo Pichini, Rodolfo Russo, Ali Seraj and Justin Vines for many very useful discussions.  We thank Gabriele Veneziano for a critical reading of the first version of our paper.
The research of FA (PDV) is fully (partially) supported by the Knut and Alice Wallenberg Foundation under grant KAW 2018.0116. Nordita is partially supported by Nordforsk.

\appendix
\section{Expressions for $\mathbf{f}(a,b,\sigma)$}
\label{AA}
The vector $\mathbf{f}$ introduced in \eqref{F15} can be written as
\begin{align}
\label{F32a}
\mathbf{f}(a,b,\sigma)=c_o(a,b,\sigma)\mathbf{b}-c_s(a,b,\sigma)\hat{\mathbf{p}}\times\mathbf{a},
\end{align}
where
\begin{align}
\label{F32a.1}
&c_o(a,b,\sigma)=\frac{b[(\sigma+\sqrt{\sigma^2-1})^2\left|\mathbf{b}-\hat{\mathbf{p}}\times\mathbf{a}\right|^2+(\sigma-\sqrt{\sigma^2-1})^2\left|\mathbf{b}+\hat{\mathbf{p}}\times\mathbf{a}\right|^2]}{2(2\sigma^2-1)\left|\mathbf{b}-\hat{\mathbf{p}}\times\mathbf{a}\right|^2\left|\mathbf{b}+\hat{\mathbf{p}}\times\mathbf{a}\right|^2},\\&c_s(a,b,\sigma)=\frac{(\sigma-\sqrt{\sigma^2-1})^2\left|\mathbf{b}+\hat{\mathbf{p}}\times\mathbf{a}\right|^2]-b[(\sigma+\sqrt{\sigma^2-1})^2\left|\mathbf{b}-\hat{\mathbf{p}}\times\mathbf{a}\right|^2}{2(2\sigma^2-1)\left|\mathbf{b}-\hat{\mathbf{p}}\times\mathbf{a}\right|^2\left|\mathbf{b}+\hat{\mathbf{p}}\times\mathbf{a}\right|^2}.
\end{align}
The components of $\mathbf{f}$ are explicitly given by
\begin{align}
&f^1(a,b,\sigma)=\frac{\left(1+\frac{(a^1)^2+(a^2)^2}{b^2}\right)\left(1-\frac{2\sigma \sqrt{\sigma^2-1}}{2\sigma^2-1}\frac{a^2}{b}\right)+\frac{2a^2}{b}\left(\frac{2\sigma\sqrt{\sigma^2-1}}{2\sigma^2-1}-\frac{a^2}{b}\right)}{(1-\frac{2a^2}{b}+\frac{(a^1)^2+(a^2)^2}{b^2})(1+\frac{2a^2}{b}+\frac{(a^1)^2+(a^2)^2}{b^2})},\\
&f^2(a,b,\sigma)=\frac{\frac{2\sigma\sqrt{\sigma^2-1}}{2\sigma^2-1}\frac{a^1}{b}\left(1+\frac{2\sigma^2-1}{\sigma\sqrt{\sigma^2-1}}\frac{a^2}{b}+\frac{(a^1)^2+(a^2)^2}{b^2}\right)}{(1-\frac{2a^2}{b}+\frac{(a^1)^2+(a^2)^2}{b^2})(1+\frac{2a^2}{b}+\frac{(a^1)^2+(a^2)^2}{b^2})},\\
& f^3(a,b,\sigma)=0.
\end{align} 
Therefore we have
\begin{align}
\label{F33a}
\mathbf{f}^2(a,b,\sigma)=\frac{1+\frac{4\sigma\sqrt{\sigma^2-1}}{2\sigma^2-1}\frac{a^2}{b}+\frac{4\sigma^2(\sigma^2-1)}{(2\sigma^2-1)^2}\frac{(a^1)^2+(a^2)^2}{b^2}}{(1-\frac{2a^2}{b}+\frac{(a^1)^2+(a^2)^2}{b^2})(1+\frac{2a^2}{b}+\frac{(a^1)^2+(a^2)^2}{b^2})}.
\end{align}

\bibliographystyle{utphys} 
\bibliography{hie-4mod.bib}

\end{document}